\documentclass[12pt,a4paper]{conference}

\usepackage{fancyhdr}
\usepackage{graphicx,amsmath,amssymb,cite}
\usepackage{multind}
\makeindex{author} \makeindex{subject}

\newcommand{\beq}{\begin{equation}}
\newcommand{\eeq}[1]{\label{#1}\end{equation}}
\newcommand{\eeqn}{\end{equation}}

\newcommand{\beqa}{\begin{eqnarray}}
\newcommand{\eeqa}[1]{\label{#1}\end{eqnarray}}
\newcommand{\eeqan}{\end{eqnarray}}

\let\bar=\overbar

\newcommand{\Dslash}{\not{\hbox{\kern-4pt $D$}}}
\newcommand{\dslash}{\not{\hbox{\kern-2pt $\del$}}}

\newcommand{\msb}{{\bar{\ssstyle M \kern -1pt S}}}

\begin{document}
%%%%%%%%%%%%%%%%%%%%%%%%%%%%%%%%%%%%%%%%%%%%%%%%%%%%%%%%%%%%%%%%%%%%%%%

\Chapter{Scalar Mesons:  A Chiral Lagrangian Framework for their Mixing 
and Substructure}
           {Scalar Mesons Mixing and Substructure}{Amir H. Fariborz}
\vspace{-6 cm}\includegraphics[width=6 cm]{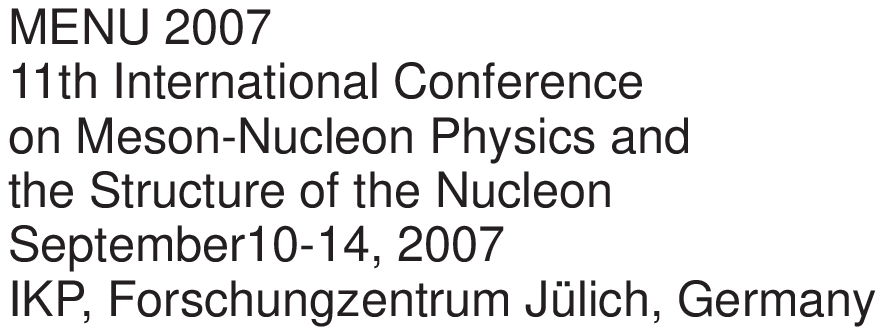}
%\bigskip\bigskip
\vspace{4 cm}

\addcontentsline{toc}{chapter}{{\it Amir H. Fariborz}} \label{authorStart}
%%%%%%%%%%%%%%%%%%%%%%%%%%%% NEW SWITCHES %%%%%%%%%%%%%%%%%%%%%%%%%%%%%%

\begin{raggedright}

{\it Amir H. Fariborz }\index{author}{Author, N.}\\
Department of Mathematics/Physics\\
State University of New York  Institute of Technology\\
 Utica, NY 13504-3050, U.S.A.
\bigskip\bigskip

%%%%%%%%%%%%%%%%%%%%%%%%%%%%%%%%%%%%%%
%%%%%%%%%%%
%%%%%%%%%%%  Repeat for second author
%%%%%%%%%%%
%%%%%%%%%%%%%%%%%%%%%%%%%%%%%%%%%%%%%%
\end{raggedright}

\begin{center}
\textbf{Abstract}
\end{center}

The highlights of studies of mixing among scalar mesons below and above 1 
GeV within a
nonlinear chiral Lagrangian framework is briefly presented.  Two scalar 
meson
nonets are introduced to explore the mass spectrum and decay properties
of the $I$=1/2 and $I$=1 scalar states.  For the $I$=0 states, in addition 
to
these two nonets a scalar glueball component is also taken into account,
and together with the constraints from the $I$=1/2 and $I$=1 sectors, 
their
mass spectrum is studied. The fact that an ideally mixed $q {\bar q}$
scalar nonet has a mass ordering which is opposite to that of an ideally 
mixed 
four-quark scalar nonet is exploited to gain some insight into the quark
substructure of the $I$=1/2, $I$=1 and $I$=0 states below and above 1 GeV.  
Consequently, numerical estimates of various components of these states
(two quark and four quark components of $I$=1/2 and $I$=1 states, and two
quark, four quark and glue component of $I$=0 states)  are determined.\\

%%%%%%%%%%%%%%%%%%%%%%%%%%%%%%%%%%%%%%%%%%%%%%%%%%%%%%%%%%%%%%%%%%%%%%%%%%

\begin{figure}[h] 
\begin{center} \includegraphics[width=13.5 cm]
{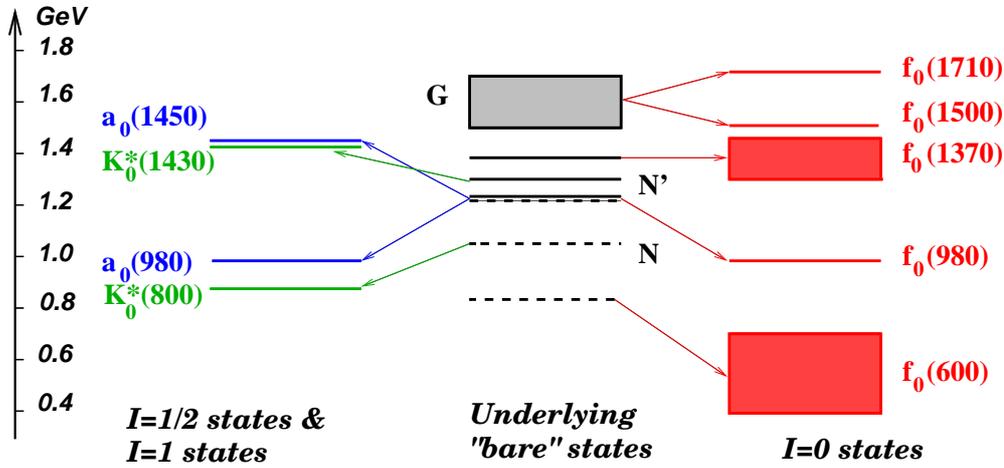} 
%\hspace{4 cm} 
\caption{ Prediction of the
present model for the substructure of the $I=1/2$, $I=1$ scalar states
below 2 GeV (left) and for the $I=0$ scalar states below 2 GeV (right) in
terms of the underlying ``bare"  states (middle).  } 
\label{F_masses}
\end{center} \end{figure}

\begin{figure}[h] 
\begin{center} 
\includegraphics[width=4. cm]{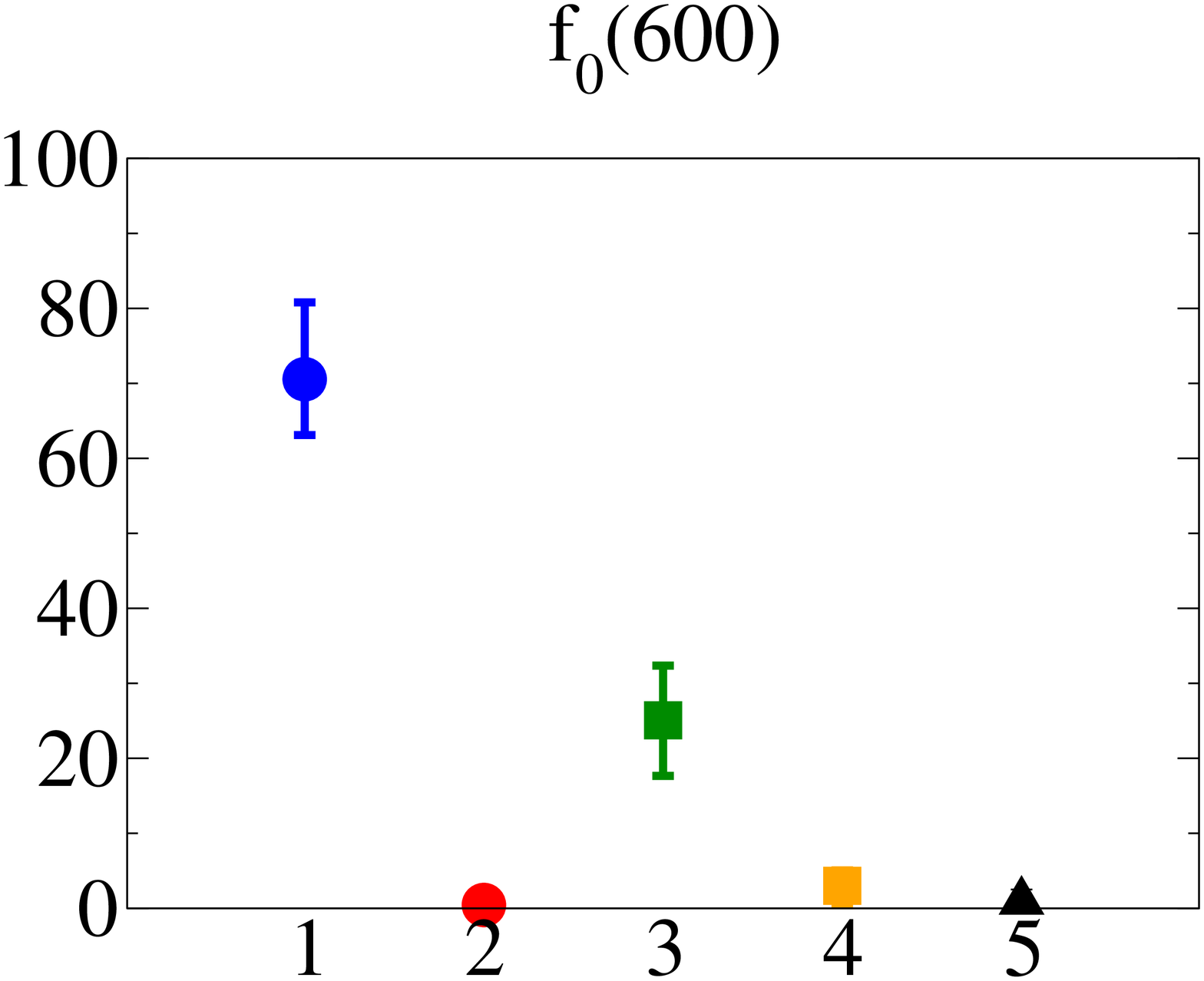} 
%\hspace{4 cm} 
\includegraphics[width=4. cm]{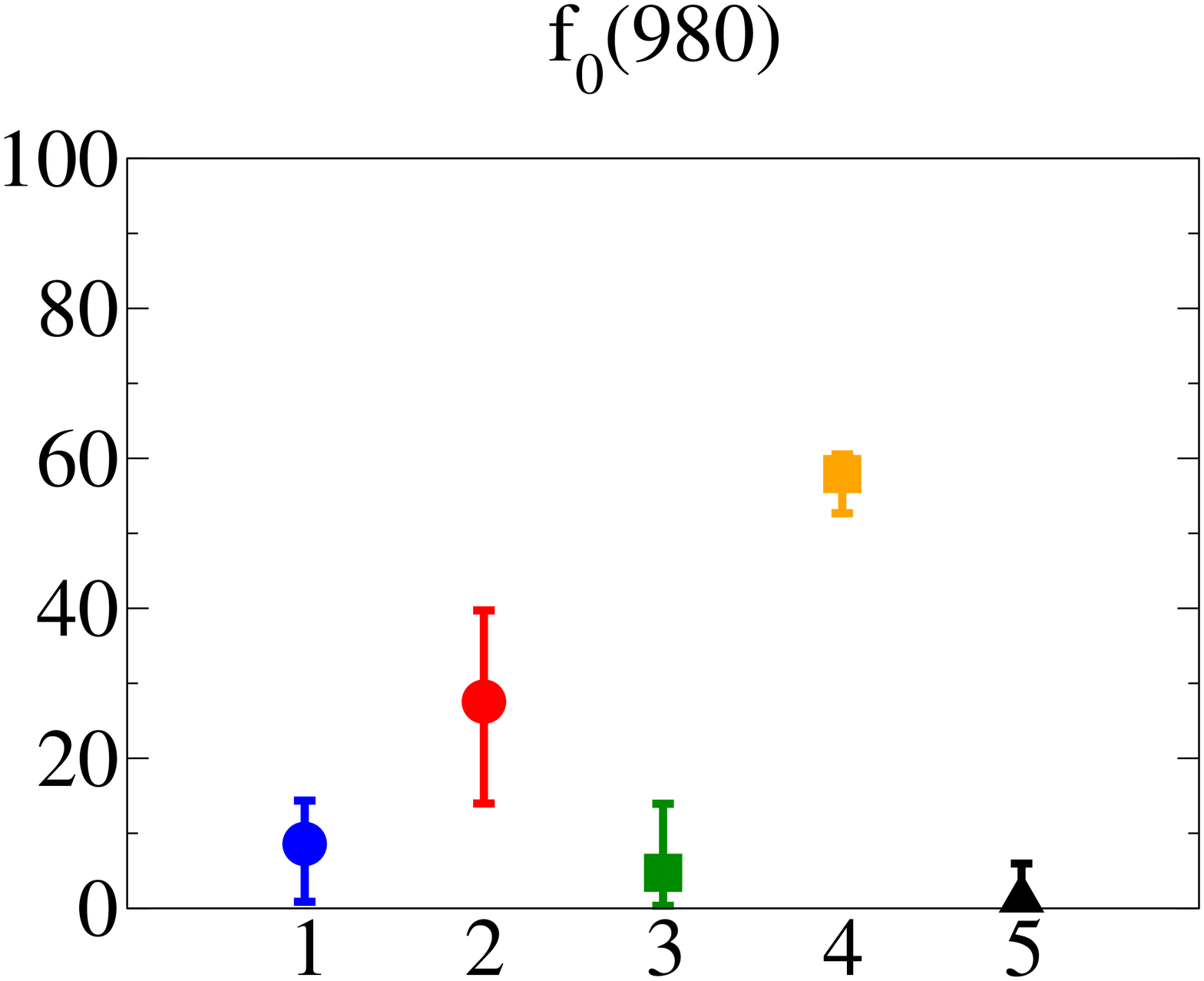} 
\includegraphics[width=4. cm]{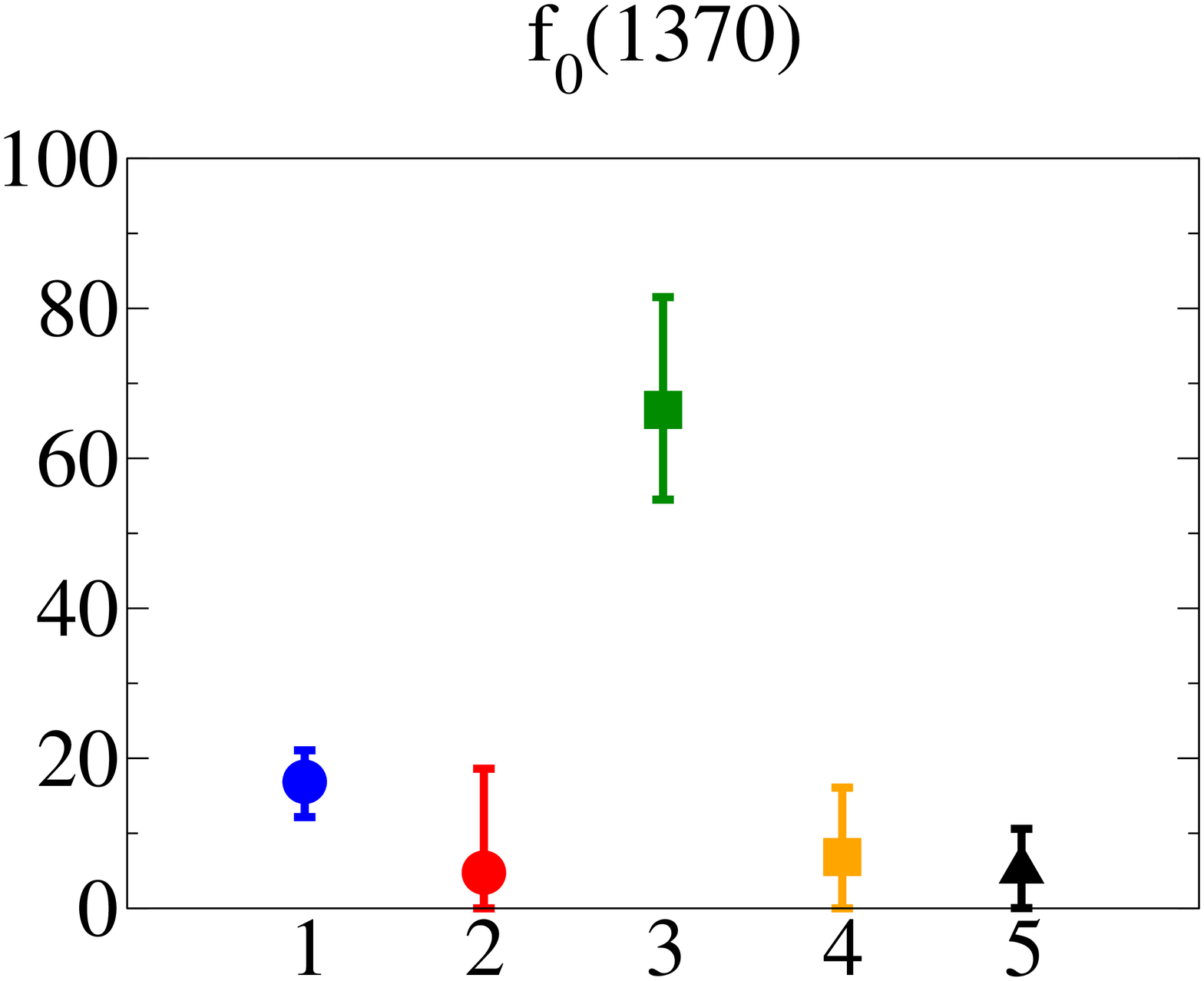}

\includegraphics[width=4. cm]{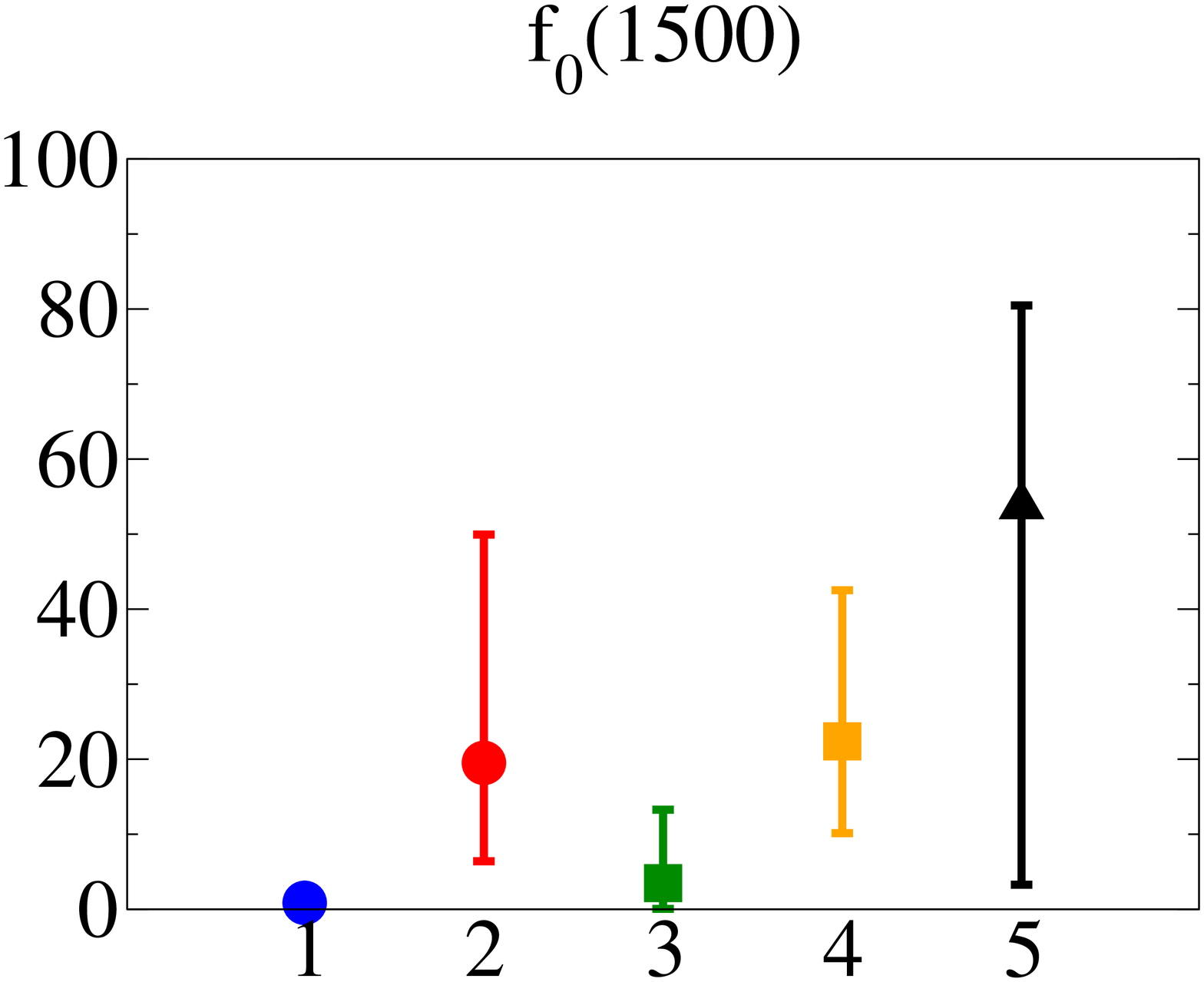}
\includegraphics[width=4. cm]{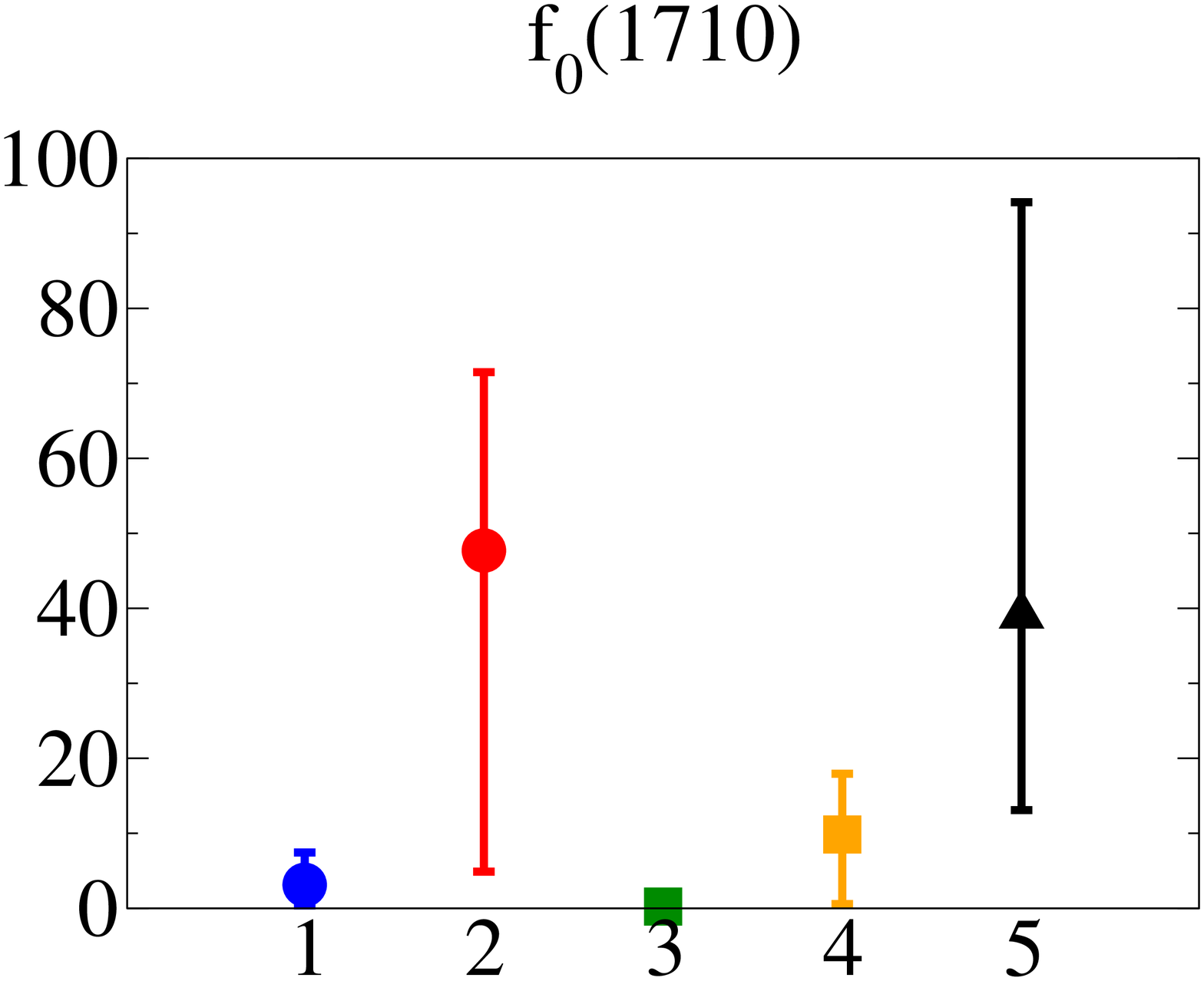}
\caption{
Components 1 to 5 respectively represent
${\bar u}{\bar d} u d$, $({\bar s}{\bar d} d s + {\bar s}{\bar u} u s)
/\sqrt{2}$, ${\bar s}s$,
$({\bar u} u + {\bar d} d)/\sqrt{2}$, and
glueball.     The symbols represent the averaged values of each component 
and
the error bars reflect the uncertainties of $m^{\exp.} [f_0(600)]$ and
$m^{\rm exp.} [f_0(1370)]$.   
} 
\label{F_mixing}
\end{center}
\end{figure}

Scalar states below and above 1 GeV are shown in Fig. \ref{F_masses}, and
are all listed/discussed in PDG \cite{PDG}.  Not all of these states are
well-established:  Among these the $f_0(600)$ [or $\sigma$] and the
$f_0(1370)$ have large uncertainties on their mass and decay widths, as
well as the $K_0^*(800)$ [or $\kappa$] which has been particularly under a
special scrutiny and debate.  It is now generally believed that the states
below 1 GeV are something other than pure $q {\bar q}$ states, as opposed
to those above 1 GeV which have been the favored candidates for a $q {\bar
q}$ nonet, even though some of their properties do not quite follow a $q
{\bar q}$ assignment.  Possible solutions for the status of the
lowest-lying scalar states include the MIT bag model, $K {\bar
K}$ molecule  and unitarized quark model, as well as
many recent investigations (see \cite{scalar_refs} for a selection of 
refs.). There are reasons to
investigate the mixing between the scalar mesons below and above 1 GeV.
First, intuitively this is not inconceivable as some of these states [such as
$f_0(600)$ and $f_0(1370)$ as well as $K_0^*(800)$] are broad and their
masses spread over a wide range, therefore one may expect that some of 
their properties may overlap. Second, the available experimental data may already be pointing to such
mixings. For example, a close look at some of the properties of the
$a_0(1450)$ and $K_0^*(1430)$ [which are expected to be two members of the
same $q {\bar q}$ scalar meson nonet (see PDG \cite{PDG})] shows
surprising deviations from a $q {\bar q}$ nonet properties.  Clearly, 
their masses are rather puzzling \cite{PDG}: If these two states belong to 
the same $q
{\bar q}$ nonet, then why should $a_0(1450)$ (which does not contain a
strange quark)  be heavier than $K_0^*(1430)$ (which does contain a
strange quark)?  There are also decay properties of these states that
cannot be understood based on a pure $q{\bar q}$ picture.  As a possible
solution, a description of the $I=1/2$ and $I=1$ scalar states below and
above 1 GeV in terms of two nonets of scalars and within a nonlinear
chiral Lagrangian framework was explored in ref.  \cite{Mec}.  In that
work, it was shown that if an underlying ``bare"  four-quark nonet $N$
lies beneath an underlying ``bare'' two-quark nonet $N'$, then as a result
of mixing of $N$ and $N'$ we can easily understand why $a_0(1450)$
becomes heavier than $K_0^*(1430)$ (in addition, the decay properties of
these states can be understood in this scenario).  Fig.  \ref{F_masses}
shows how this mechanism works.  It was also found in \cite{Mec} that the
$I=1$ states are close to equal admixtures of two and four-quark states,
whereas the $I=1/2$ states are less mixed, with $K_0^*(800)$ containing
close to 75\% four-quak and 25\% two-quark [and vice versa for
$K_0^*(1430)$]. What does this scenario say about the $I=0$ states?  This
question was studied in \cite{Far} in which the implications
of such underlying mixing of nonets $N$ and $N'$ on the $I=0$ states was
investigated.  Fig.  \ref{F_masses} summarizes the results and shows how
the $I=0$ states originate from the four-quark nonet $N$, two-quark nonet
$N'$ and a scalar glueball $G$.  The mass part of the Lagrangian for $N$, 
$N'$ and $G$ is (in the leading order of mixing): 
\begin{eqnarray} 
&& \hskip - .8cm {\cal L}_{mass}=
- a {\rm Tr}(NN) - b {\rm Tr}(NN{\cal M})  - a' {\rm Tr}(N'N') - b' {\rm
Tr}(N'N'{\cal M})  
\nonumber \\ 
&& \hskip - .5cm   
- c {\rm Tr}(N){\rm Tr}(N)  - d {\rm Tr}(N) {\rm Tr}(N{\cal M})  - c' {\rm
Tr}(N'){\rm Tr}(N')  
- d' {\rm Tr}(N') {\rm Tr}(N'{\cal M})  
\nonumber \\ 
&&  \hskip - .5cm  
-\gamma {\rm Tr} \left( N N' \right)  
- \rho {\rm Tr} (N) {\rm Tr} (N') - g G^2 
- e G {\rm Tr} \left( N
\right) - f G {\rm Tr} \left( N' \right) 
\label{L_mix_I0} 
\end{eqnarray}
in which ${\cal M}$ is the usual quark mass spurion.  The mass of the
$I=1/2$ and $I=1$ states involve terms $a$, $b$, $a'$, $b'$ and $\gamma$
only.  The mass of $I=0$ states involve all 13 parameters.  The mixing of
$I=0$ states is clearly much more complicated and amounts to 5$\times$5
rotation matrices among $N$, $N'$ and $G$.  The result of the numerical
analysis of  \cite{Far} for the prediction of the substructure of the
$I=0$ states are given in Fig. \ref{F_mixing}, in which, in the middle,
the dashed lines represent nonet $N$ (that has a mass 
ordering
consistent with an ideally mixed four-quark nonet), the solid
lines represent nonet $N'$ (that has a mass ordering consistent with an
ideally mixed two-quark nonet) and the box represents the scalar glueball
predicted in this model.  Identifying the components of the two bare
nonets with the corresponding members of an ideally mixed four-quark nonet
and ideally mixed two-quark nonet results in conclusion that the bare
masses in nonet $N$ are (from bottom to top): $m({\bar u}{\bar d} u d)$ =
0.83 GeV, $m({\bar d}{\bar s} u d)$ = 1.06 GeV, $m\left[({\bar s}{\bar d}
d s + {\bar s}{\bar u} u s) /\sqrt{2}\right]$ = 1.24 GeV; and the bare
masses in nonet $N'$ are (from bottom to top): $m\left[({\bar u} u + {\bar
d} d)/\sqrt{2}\right]$ = 1.24 GeV, $m({\bar u} s)$ = 1.31 GeV and $m({\bar
s}s)$ = 1.38 GeV.  The uncertainty of the glueball mass (shown by the
height of the box, approximately between 1.5 GeV to 1.7 GeV) is due to the
uncertainty of the input masses of $f_0(600)$ and $f_0(1370)$.  In Fig.
\ref{F_masses}, on the right, the $I=0$ physical states are shown, and the
height of the two boxes represent the prediction of the present model for
the uncertainties of the masses of $f_0(600)$ and $f_0(1370)$, which are 
(in
this model) approximately in ranges $0.4-0.7$ GeV and $1.3-1.45$ GeV,
respectively.  On the left, the $I=1/2$ and $I=1$ physical states are
shown [note the level-crossing that explains the properties of $a_0(1450)$
and $K_0^*(1430)$].  The arrows show the dominant component of each
physical state.  Finally, the detailed numerical analysis of \cite{Far}
predicts the substructure of the $I=0$ scalars (in terms of two quark,  
four quark and  glueball components) which are given in Fig.  
\ref{F_mixing}.
%%%%%%%%%%%%%%%%%%%%%%%%%%%%%%%%%%%%%%%%%%%%%%%%%%%%%%%%%%%%%%%%%%%%%%%%%%%%5
%%%%%%%%%%%%%%%%%%%%%%%%%%%%%%%%%%%%%%%%%%%%%%%%%%%%%%%%%%%%%%%%%%%%%%%%%%%%%
%%%%%%%%%%%%%%%%%%%%%%%%%%%%%%%%%%%%%%%%%%%%%%%%%%%%%%%%%%%%%%%%%%%%%%%%%%%%5
%%%%%%%%%%%%%%%%%%%%%%%%%%%%%%%%%%%%%%%%%%%%%%%%%%%%%%%%%%%%%%%%%%%%%%%%%%%%%
\section*{Acknowledgments}
The author wishes to thank the organizers of MENU 2007 in particular Dr. 
S. Schadmand for her help at various stages of the conference.
This work has been supported by the NSF Award 0652853.
%\begin{thebibliography}{000} %for 3 digits
%\begin{thebibliography}{00}  %for 2 digits

%%%%%%%%%%%%%%%   Author and Subject Index
%\printindex{author}{Author Index}
%\blankpage

%\printindex{subject}{Subject Index}
% \blankpage
\end{document}